\documentstyle[emulateapj]{article}

\submitted{Received 1998 October 30; accepted 1998 December 28}

\received{30 October 1998}
\accepted{28 December 1998}

\lefthead{ZUBKO}
\righthead{ON THE MODEL OF DUST IN THE SMALL MAGELLANIC CLOUD}

\begin{document}

\title{On the model of dust in the Small Magellanic Cloud}

\author{Victor G. Zubko\altaffilmark{1}}
\affil{Department of Physics, Technion -- Israel Institute
             of Technology, Haifa 32000, Israel; zubko@phquasar.technion.ac.il}

\altaffiltext{1}{On leave from the Main Astronomical Observatory,
                 National Academy of Sciences of the Ukraine, Kiev}

\begin{abstract}
We present here dust models for the Small Magellanic Cloud bar calculated
for the first time with the regularization approach. A simple mixture
of the core--mantle and/or composite grains (mostly made from silicates
and organic refractory) together with silicon nanoparticles is consistent
with the following: 1) the observed extinction toward AzV 398 (a typical SMC
bar sightline), 2)~the~elemental abundances, and 3) the strength of
the interstellar radiation field. We predict the expected albedo and asymmetry parameter
of the models, which can be tested in future observations.
The proposed dust models can also be tested by looking for the expected
extended red emission.
\end{abstract}

\keywords{dust, extinction -- galaxies: individual (Small Magellanic Cloud) --
galaxies: ISM -- Magellanic Clouds
}

\section{Introduction}

The Small Magellanic Cloud (SMC), one of our nearest neighbours,
contains considerably less heavy elements and dust than the Galaxy
(Bouchet et al. \cite{bouchet}; Welty et al. \cite{welty}).
In addition, dust in the SMC  seems to be different
from dust in the Galaxy (Pr\'evot et al. \cite{prevot}; Pei \cite{pei}).
For example, a typical extinction curve of the SMC is almost linear
with inverse wavelength and does not show a presence of the UV bump
at 2175 {\AA} (Pr\'evot et al. \cite{prevot}; Thompson et al. \cite{thomp}).
Recently, Rodrigues et al. (\cite{rodr}) have measured the linear polarization
for a sample of the SMC stars in the optical and found that
the wavelength of maximum polarization is generally smaller than
that in the Galaxy. The low metallicity of the SMC implies that
it should be at an early stage of its chemical evolution, thus resembling 
in this respect galaxies at high redshifts. A strong support to this view
has come from the discovery that the dust in starburst galaxies,
apparently the only type of galaxies found so far for redshifts $z>2.5$,
has an extinction law remarkabely similar to that in the SMC
(Gordon, Calzetti, \& Witt \cite{gcw}).

Several attempts have been made to model the dust in the SMC.
Bromage \& Nandy (\cite{bn83}) and Pei (\cite{pei}) modelled
some SMC mean extinction curves using the dust mixture of spherical
graphite and silicate grains with a power-law size distribution
(Mathis, Rumpl \& Nordsieck \cite{mrn}, hereafter MRN;
Draine \& Lee \cite{dl84}). Bromage \& Nandy (\cite{bn83}) and Pei
(\cite{pei}) have shown that an MRN-like mixture with a lower fractional
mass of graphite grains in comparison to silicate grains, and with other
parameters as in the Galaxy, can satisfactorily explain the SMC extinction.
Pei (\cite{pei}) even succeeded to fit the SMC extinction law with
silicate grains alone. Rodrigues et al. (\cite{rodr}) made model fits
to both extinction and polarization for two stars in the SMC, AzV 398 and
AzV 456, representing lines of sight with different properties.
The authors have found that the mixture of bare silicate and amorphous
carbon, or graphite spheres together with silicate cylinders with an MRN-like
power-law size distribution explains quite well both the extinction and
the polarization.

However, the SMC dust models proposed so far are too simplistic concerning
both the choice of grain constituents and grain-size distributions.
Recently, Mathis (\cite{m96}), Li \& Greenberg (\cite{lg97}) and Zubko,
Kre{\l}owski, \& Wegner (\cite{zkw2}) have presented more sophisticated
models of Galactic dust using core-mantle, multilayer and composite grains
which are much more physically justified then the models previously considered.
Note that the models by Zubko et al. (\cite{zkw2}) were calculated
with the regularization approach (Zubko \cite{zubko}) which is
a very efficient method capable of deriving optimum and {\em unique}
size distributions in a {\em general} form for any predefined mixture of
grains by simultaneously fitting the extinction curve, the elemental
abundances and the mass fraction constraints. The uniqueness of the grain
size distributions follows from the mathematical nature of the problem when
we need to solve a Fredholm integral equation of the first order, being
a typical ill-posed problem. The regularization approach reduces this problem
to minimization of a strongly convex quadratic functional. The latter
problem was strictly proved to have a {\em single} solution.
See for more details, e.g. Groetsch (\cite{groe84}), Tikhonov et al.
(\cite{tikh90}) or Zubko (\cite{zubko}). This method can be expanded
to allow one to deduce the uncertainty in the solution
based on the data uncertainty, but this is beyond the scope of
this letter (Zubko 1999, in preparation).

Recently, Gordon \& Clayton (\cite{gc98}) have derived the extinction curves
for four SMC stars with several improvements in comparison to previous
studies: higher S/N {\em IUE} spectra, and a more careful choice of
the pairs of reddened and comparison stars. The sightlines toward three
stars: AzV 18, 214, and 398, located in the SMC bar pass through the regions
of active star formation and exhibit similar extinction law.
The sightline toward the star AzV 456, located in the SMC wing,
passes through a much more quiescent region of star formation and shows
a Galaxy-like extinction with the 2175 {\AA} UV bump. The purpose of
the present Letter is to report the first SMC dust models calculated
with the regularization approach to the new high quality extinction curves.
We modelled the extinction toward the star AzV 398 which is thought to be
a typical SMC bar sightline (Rodrigues et al. \cite{rodr}; Gordon \& Clayton
\cite{gc98}). The results of the study which includes all the four stars
will be presented in a forthcoming paper.

\section{Empirical data}
     \label{sec:emp_data}

We transform the extinction curve for AzV 398, derived by Gordon \& Clayton
(\cite{gc98}) in the standard form:
$E(\lambda)=[\tau(\lambda)-\tau(V)]/[\tau(B)-\tau(V)]$,
to the extinction cross section per H atom:
\begin{equation}
{ \tau(\lambda) \over N_{\rm{H}} }= 0.921 {E(\bv) \over N_{\rm{H}} }
  [E(\lambda)+R_V]
\end{equation}
where $\tau$ is the optical thickness, $E(\bv)$ is the \bv colour excess,
$R_V$ is the total-to-selective extinction ratio, and $N_{\rm{H}}$ is
the column number density of hydrogen. For $N_{\rm{H}}$ we take the value
1.5$\times$10$^{22}$ cm$^{-2}$ from Bouchet et al. (\cite{bouchet}),
corresponding to atomic hydrogen. Since the sightline toward AzV 398 is
associated with an \ion{H}{2} region (Gordon \& Clayton \cite{gc98}),
it is likely that the contribution of molecular hydrogen to $N_{\rm{H}}$
is neglegibly small (for sightlines not passing through the SMC bar this
may not be the case, see Lequeux \cite{lequeux}). The value for $E(\bv)$=0.37
was also taken from Bouchet et al. (\cite{bouchet}) and $R_V$=2.87 from
Gordon \& Clayton (\cite{gc98}).

The elemental abundances (gas + dust), currently adopted for the SMC, were
taken from Welty et al. (\cite{welty}): C/H=46 p.p.m. (atoms per 10$^6$ H
atoms) or 7.66$\pm$0.13 dex, O/H=107 p.p.m. or 8.03$\pm$0.10 dex,
Si/H=10 p.p.m. or 7.00$\pm$0.18 dex, Mg/H=9.1 p.p.m. or 6.96$\pm$0.12 dex,
and Fe/H=6.6 p.p.m. or 6.82$\pm$0.13 dex. Note that these values
are 2--5 times less than the respective Galactic abundances following
recent revision (Snow \& Witt \cite{sw}; Cardelli et al. \cite{cardelli}).
Since we have no information on the amounts of elements in dust and gas
separately in the SMC, we simply assume in this study that as in the Galaxy
42\% of carbon ($\sim$20 p.p.m.), 37\% of oxygen ($\sim$40 p.p.m.), and
all silicon, magnesium and iron are locked up in dust (Cardelli et al.
\cite{cardelli}; Zubko et al. \cite{zkw2}). The actual amount of
elements locked up in dust is uncertain, but one may expect even lower
amounts of elements in dust because the SMC is less chemically processed
than our Galaxy. 

Recently, Witt, Gordon, \& Furton (\cite{wgf}) and Ledoux et al. (\cite{ledoux})
proposed that the silicon nanoparticles might be the source of
the extended red emission (ERE) in our Galaxy. Zubko, Smith, \& Witt
(\cite{zsw}) have modeled the mean Galactic extinction curve with
the silicon nanoparticles involved and have shown that this hypothesis is
consistent with the available data on extinction, elemental abundances and
the ERE. On the other hand, Perrin, Darbon, \& Sivan (\cite{pds})
and Darbon, Perrin, \& Sivan (\cite{dps}) have revealed ERE in
extragalactic objects showing active star formation: the starburst galaxy M82
and the nebula 30 Doradus in the Large Magellanic Cloud, respectively.
Since the sightline toward AzV 398 passes through a star-forming region,
we may expect to observe ERE from there as well. We thus included the silicon
nanoparticles in our modeling. As in Zubko et al. (\cite{zsw}), we used
the silicon core--SiO$_2$ mantle model of silicon nanoparticles with
optical constants of nanosized silicon from Koshida et al. (\cite{koshida}).
We also included in present study the grain constituents
(graphite, silicate, SiC, organic refractory, amorphous carbon, water ice
and others) and respective optical constants previously used by
Zubko et al. (\cite{zkw2}).

\section{Models of extinction}
       \label{sec:mod_ext}

We performed extensive work on modeling the extinction curve for AzV 398,
searching for the physically reasonable mixtures of dust constituents.
Our goal was to find the models which would simultaneously fit the extinction,
consume the allowed amounts of chemical elements and include silicon
nanoparticless (by analogy with the Galactic case, Zubko et al. \cite{zsw}).
We report in this Letter three simple models which fulfil all the above
requirements. The results are presented in
Figs~\ref{fig:models}--\ref{fig:scat_prop} and Table~\ref{tab:tab1}.
The model grains are mostly made up of two species: silicate (MgFeSiO$_4$)
and organic refractory residue, which coexist either in
core(silicate)-mantle(organic refractory) or in spherical porous composite
grains with the latter containing also small amounts of amorphous carbon.
The total mass fraction of silicate + organic refractory is about 0.9.
The other important model component is the silicon nanoparticles,
which are found to have a mass fraction of about 0.07--0.085.

The fact that organic refractory is among the major grain constituents is in
good agreement with the expectations that the interstellar radiation field
(ISRF) in the SMC is stronger than in the Galaxy (Lequeux et al.
\cite{leqetal}) since the icy mantles on silicate grains formed in molecular
clouds can be processed by the UV radiation into organic refractory
(Greenberg \& Li \cite{gl96}). Note especially that our attempts
to include silicate and SiC grains coated by either amorphous carbon
or water ice mantles and also bare carbonaceous (graphite, amorphous
carbon), silicate and SiC grains resulted in very low mass fractions
of such grains, typically less than 1 per cent. This means that
the conditions in the SMC (ISRF intensity, duration of the exposure by
the UV radiation) are probably favourable for converting icy mantles into
organic refractory, but not for the further processing of organic
refractory into amorphous carbon. In contrast to the SMC, the presense of
icy mantles may be allowed for the dust grains in the Galactic diffuse
medium (Zubko et al. \cite{zkw2}).

\placefigure{fig:models}

Following Zubko et al. (\cite{zkw2}), the models based on the silicate
core--organic refractory mantle (composite) grains are refered to as G
(M) models. The GM model is a combination of G and M models and contains both
core-mantle and composite grains. As shown in Fig.~\ref{fig:models}
all the above models fit the extinction curve quite well.
The size distributions of both core-mantle and composite grains
are quite wide and cover both small and larger grains with the preference
to the grains of sizes 10--100 nm. Silicon nanoparticles have a diameter
of 3.0 nm by definition. All the models consume the maximum amounts of carbon,
oxygen and silicon allowed for dust and slightly less for magnesium and iron.
All the carbon consumed is contained in organic refractory. Approximately
equal amounts of silicon are locked up in the silicon nanoparticles
(4--5 p.p.m.) and in other components (5--6 p.p.m.). Silicate core-organic
refractory mantle grains prevail by mass in all the cases.
Note that the dust composition in our models is drastically different from
that in previous SMC models, which were based on modification of the standard
MRN model (Pei \cite{pei}; Rodrigues et al. \cite{rodr}).
The grain size distributions in our models are not a simple power law,
as assumed in the previous models, but are rather optimized to reproduce
the extinction curve and to simultaneously obey the abundance constraints.

\placefigure{fig:scat_prop}

\placetable{tab:tab1}

Since we do not know presently the wavelength-dependent intensity
of the ISRF in the SMC and, in addition, the existing extinction curves 
for the SMC have a UV boundary at around 0.13 $\mu$m, we are unable
to estimate the fraction of the UV photons absorbed by the silicon
nanoparticles. In the Galaxy, Gordon et al. (\cite{gwf}) found this fraction
to be 0.10$\pm$0.03. However, as was found by Zubko et al. (\cite{zsw}),
the mass fraction of silicon nanoparticles may serve as a good indicator
in this case. The values of this quantity derived for the models presented
above: 0.07, 0.071, and 0.085 suggest that the SMC is similar to the Galaxy,
and therefore should also be a source of significant ERE.
Moreover, the proximity of the silicon nanoparticle mass fractions obtained
by modeling rather different extinction laws: Galactic and SMC,
with different chemical constraints and different dust models suggests
that silicon nanoparticles and ERE may be a universal phenomena in galaxies.

It is evident from Fig.~\ref{fig:models} and Table~\ref{tab:tab1}
that each of the models presented is almost equally good (as indicated
by $\bar \mu$) in fulfilling the requirements formulated above, with the G
model being slightly more preferable. In order to discriminate between
the models, we calculated the model scattering properties, albedo
and asymmetry parameter, which are displayed in Fig. \ref{fig:scat_prop}.
We compare our results with the observational data for the Galaxy
taken from Gordon et al. (\cite{gcw}) to illustrate the significantly
different predictions for the SMC. The model albedos are quite close
to one other for all wavelengths, whereas the model asymmetry parameter
shows large differences, especially in the UV. In general, the model
albedos are lower than respective Galactic ones, except for the near IR,
where we may see an opposite effect. Only the asymmetry parameter of M model
is close to the respective Galactic values. Note that the albedos of
Pei's  (\cite{pei}) model of the SMC dust significantly exceed the albedos,
calculated in our models, and also the expected Galactic values
(Fig. \ref{fig:scat_prop}). Another possible mean to choose the most
appropriate model may be the polarization data. We hope to include these
data in a self-consistent analysis in forthcoming papers.

In summary, we report here for the first time more refined models of the SMC
bar dust which are in good agreement with the observed extinction, elemental
abundances, and the strength of the ISRF. The models were calculated by using
the regularization approach. The major grain constituents were found to be
silicates, organic refractory and nanosized silicon. This conclusion
is subject to some uncertainty due to the uncertain element depletion
patterns in the SMC. We predicted the scattering properties of our models,
which are significantly different from the Galactic values.
More observational constraints, e.g. the polarization data, are
to be included into analysis to choose the most appropriate dust model.

\begin{acknowledgments}
I thank Karl Gordon and Geoff Clayton for providing me the extinction
curve for AzV~398 in the electronic table.
During my work, I benefited from many stimulating discussions with Ari Laor.
This research was supported by a grant from the Israel Science Foundation.
\end{acknowledgments}

\clearpage

{\plottwo{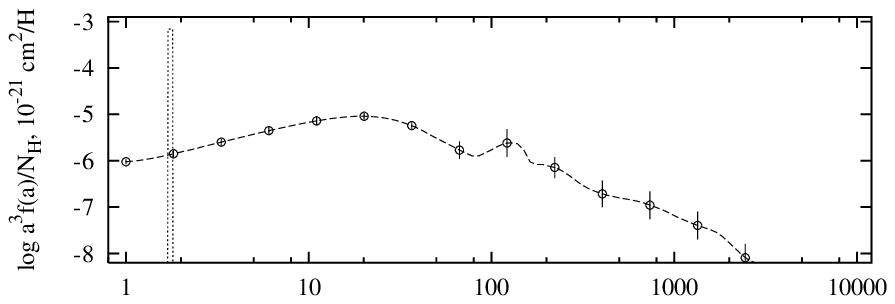}{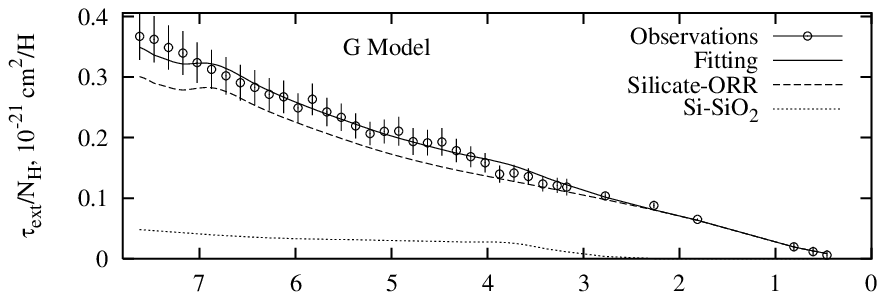}}

{\plottwo{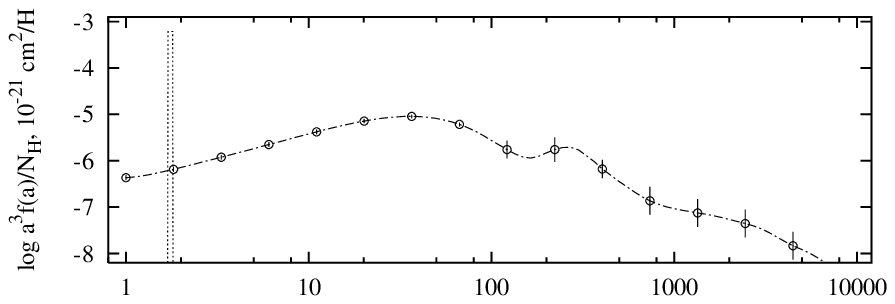}{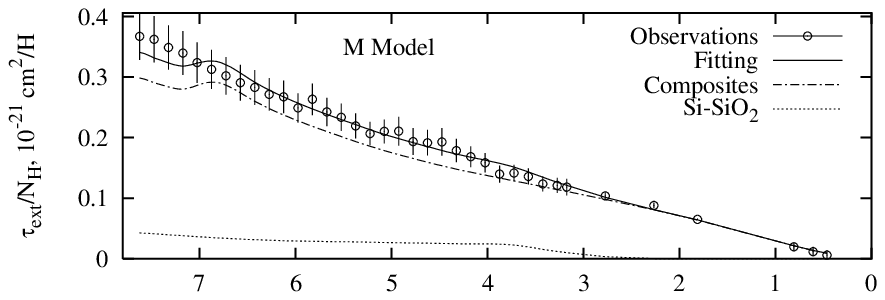}}

{\plottwo{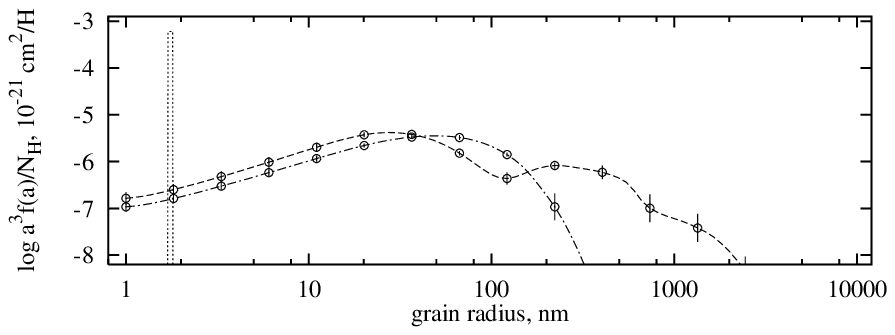}{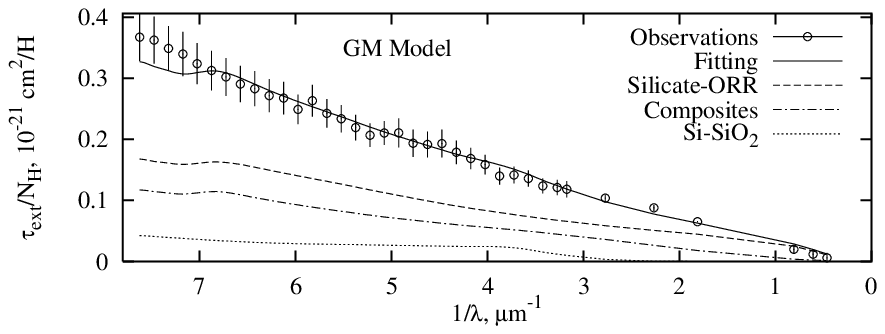}}

\figcaption[df1.eps ec1.eps df2.eps ec2.eps df3.eps ec3.eps]{
     Size distributions of dust grains (left panels) and
     respective extinction curves (right panels), fitting the extinction
     curve toward AzV 398. Shown are G, M and GM dust models.
     The uncertainties of the size distributions were calculated
     using the method of statistical modeling by Zubko (in preparation).
      \label{fig:models}
}

{\plottwo{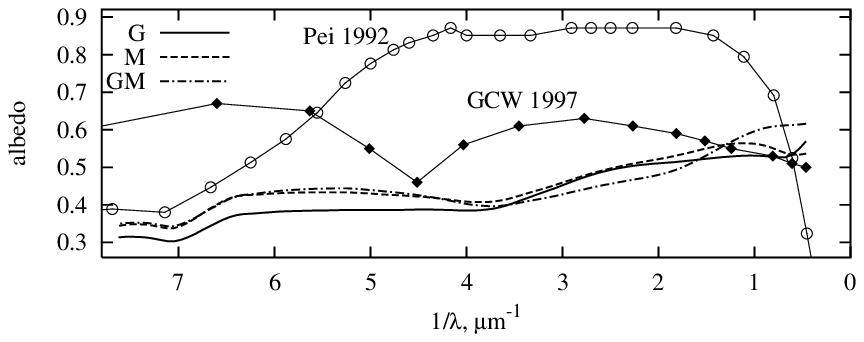}{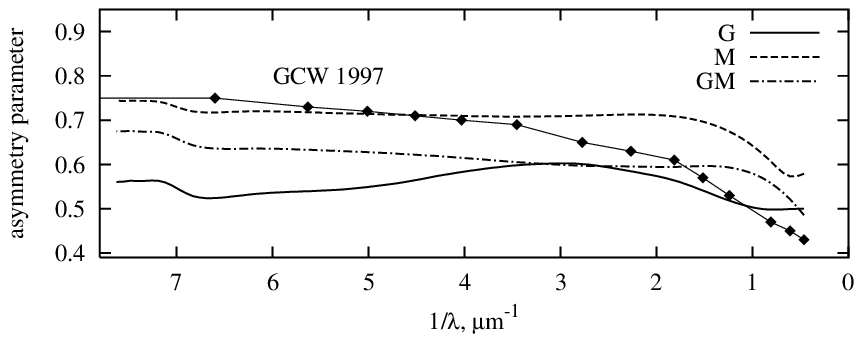}}

\figcaption[albedo.ps, aspar.ps]{
     Scattering properties: albedo and asymmetry parameter,
     corresponding to various dust models fitting the extinction
     curve toward AzV 398. The diamonds depict the observational data
     for the Galaxy taken from Gordon et al. (1997) and the circles
     the albedo of Pei's (1992) model of the SMC dust.
         \label{fig:scat_prop}
}


\begin{deluxetable}{clrrrrrrcc}
\tablefontsize{\small}
\tablecaption{The main parameters of the models fitting the extinction toward
   AzV 398. \label{tab:tab1}}
\tablehead{
  \colhead{Model } &
  \colhead{Components } &
  \colhead{C } &
  \colhead{Si } &
  \colhead{O } &
  \colhead{Mg } &
  \colhead{Fe } &
  \colhead{$f_{\rm{mass}}$ } &
  \colhead{$M$ } &
  \colhead{$\bar \mu$ }
}
\startdata
SMC   & gas + dust                 &  46 & 10 & 107 &  9 &  7 & & & \nl
      & dust                       &  20 & 10 &  40 &  9 &  7 & & & \nl\nl

  G   &                            &  20 & 10 &  40 &  5 &  5 & 100.0 & 2.62-27& 0.211 \nl

      & Silicate(0.4)-ORR(0.6)     &  20 &  5 &  40 &  5 &  5 &  91.5 &       & \nl
      & Si(0.99)-SiO$_2$(0.01)     &   0 &  5 &   0 &  0 &  0 &   8.5 &       & \nl\nl

  M   &                            &  20 & 10 &  40 &  6 &  6 & 100.0 & 2.81-27& 0.269 \nl

      & Composites (45\%-porous)   &  20 &  6 &  40 &  6 &  6 &  93.0 &       &  \nl
      & Si(0.99)-SiO$_2$(0.01)     &   0 &  4 &   0 &  0 &  0 &   7.0 &       &  \nl\nl

 GM   &                            &  20 & 10 &  40 &  6 &  6 & 100.0 & 2.75-27& 0.233 \nl

      & Silicate(0.5)-ORR(0.5)     &  13 &  5 &  31 &  5 &  5 &  70.4 &       &  \nl
      & Composites (45\%-porous)   &   7 &  1 &   9 &  1 &  1 &  22.5 &       &  \nl
      & Si(0.99)-SiO$_2$(0.01)     &   0 &  4 &   0 &  0 &  0 &   7.1 &       &  \nl\nl
\enddata

\tablenotetext{a}{
  First two lines contain the SMC elemental abundances in gas+dust and dust.
  Next lines present the data for the models reported: elemental abundances
  in ppm (atoms per 10$^6$ H atoms),
  mass fractions of each dust component ($f_{\rm{mass}}$), the total mass
  of dust matter in g per H atom ($M$) and the values of $\bar \mu$,
  characterizing a quality of a model (see Zubko et al. \cite{zkw1} for more
  details). The volume fractions of grain constituents are noted in brackets. 
  The 45\%-porous composite grains mostly contain silicate (0.28 in model M
  and 0.23 in model GM) and organic refractory residue (ORR)(0.24 and 0.23)
  with less amount of amorphous carbon of various forms (0.09 and 0.03).
}

\end{deluxetable}

\end{document}